\documentstyle[11pt,newpasp,twoside]{article}
\index{pulsars}
\index{Parkes}

\begin{document}
\title{Pulsar Birthrate from Parkes Multi-beam Survey}
\author{N. Vranesevic\altaffilmark{1}, R. N. Manchester, G. B. Hobbs, 
A. G. Lyne, M. Kramer, D. R. Lorimer, F. Camilo, I. H. Stairs, V. M. Kaspi, 
N. D'Amico, A. Possenti}
\altaffiltext{1}{University of Sydney, School of Physics, NSW 2006, Australia, \& 
ATNF--CSIRO, P.O. Box 76, Epping, NSW 1710, Australia}


\begin{abstract}
We report on calculations of the pulsar birthrate based on the 
results of the Parkes multibeam survey. From the observed sample of more 
than 800 pulsars, we compute the pulsar current, accounting as accurately as 
possible for all known selection effects. The main goal of this work is to 
understand the pulsar birthrate as a function of the
surface dipole magnetic field strengths. We show that pulsars with magnetic 
fields greater than $10^{12.5}$ G account for about half of the total 
birthrate. 
\end{abstract}

\section{Introduction}

The Parkes multi-beam pulsar survey covers a $10^\circ$-wide strip of the 
southern Galactic plane from $l=260^\circ$ to $l=50^\circ$ (Manchester et al. 
2001). It utilises a 13-beam receiver operating in the 20-cm band on the 
Parkes 64-m radio telescope and is seven times more sensitive than any 
previous large-scale survey. The survey has discovered more than 600 new 
pulsars so far, including many that are young and energetic. 
Already published pulsars can be found in the on-line catalogue 
at {\tt http://www.atnf.csiro.au/research/pulsar/psrcat}. In the following 
analysis we used 520 pulsars that were discovered by the Parkes survey, of 
which 415 are obtained from public catalogue and 105 are currently unpublished.
These new discoveries, together with 210 previously known pulsars, result
in a sample of $N_{\rm psr} = 730$ pulsars, with a mean period of 0.76 s. 

\section{Pulsar current calculation}

The observed distribution of pulsars in the Galaxy differs systematically from 
the true distribution due to various observational selection effects 
inherent in pulsar surveys (see the contribution by Lorimer in these 
proceedings for further details). We have accurately modeled the
sensitivity threshold of the multibeam survey due to these effects and
use this model to compute a weight or scale factor for each pulsar by: 
1) placing 
it at large number of randomly selected locations in the model Galaxy; 
2) calculating the effective dispersion measure and the interstellar 
scattering;
and 3) recording the number of detections, i.e. those positions for which 
the predicted flux density exceeds the survey limit. 
For each pulsar, then, the scale factor is simply the
ratio of the total number of locations in our model Galaxy 
(typically $10^5$) to the number of detections. This results in
an estimate of the number of similar pulsars in the Galaxy. 

The pulsar current, $J$, is the flow of objects from short to long spin periods.
We can compute $J$ as a function of period, $P$, by binning pulsars into intervals
of width $\Delta P$. The current in a given bin is given by:
\begin{displaymath}
J(P)=\frac{1}{\Delta P}\left(\sum_{i=1}^{n_{\rm psr}}\frac{S_i\dot{P_i}}{f}\right),
\end{displaymath}
where $S_i$ is the scale factor for each pulsar, $n_{\rm psr}$ is the number of pulsars 
in the bin and $f$ is the fraction of $4\pi$ sr covered by the emission beam.
In this analysis, we assume for simplicity that $f=0.2$. Under steady-state
conditions, the maximum value of $J$ is the birthrate of pulsars. The total 
number of active pulsars in the Galaxy is given by the sum of $S_i/f$.
For further details, see Lorimer (these proceedings) and also Vivekanand \& Narayan (1981).

\section{Results}

The results are summarized in Table 1 for both the
TC93 distance model (Taylor \& Cordes 1993) and the newer 
NE2001 model (Cordes \& Lazio 2002). Our main conclusion is that
pulsars with surface dipole magnetic fields greater than $10^{12.5}$ G account 
for half of the total birthrate. For both distance models,
the birthrate of these high-field pulsars is 52\% of the total population.
Low-field ($< 10^{12}$ G) pulsars contribute only 10\% of the
total birthrate using the TC93 model, or 17\% for NE2001.
Full details of this analysis are now being prepared for publication.

{\small
{\begin{table}
\caption{\small {Pulsar birthrate and total Galactic population for four sets 
of data: total plus three ranges of magnetic field strengths.}}
\begin{tabular}{l l c c r r}
\tableline
 & & No. in & birthrate & birth    & No. in \\  
 & & sample & pulsars/century & interval & Galaxy \\
\tableline
TC93 & total & 723 & $2.53 \pm 0.70$ &  31--55 yr & 259000 \\
& ${\rm log}(B)\geq 12.50$ & 166 & $1.32 \pm 0.58$ & 53--135 yr & 11000\\
& $12.00 \leq {\rm log}(B) < 12.50$ & 301 & $0.95 \pm 0.34$ & 78--164 yr & 
38000 \\ 
& ${\rm log}(B) < 12.00$ & 256 & $0.26 \pm 0.16$ & 238--1000 yr & 210000 \\
\tableline
NE2001 & total & 721 & $4.50 \pm 1.16$ & 18--30 yr & 660000\\
& ${\rm log}(B)\geq 12.50$ & 168 & $2.35 \pm 0.91$ & 31--69 yr & 17000\\
& $12.00 \leq {\rm log}(B) < 12.50$  & 299 & $1.40 \pm 0.49$ & 53--110 yr & 
80000\\ 
& ${\rm log}(B) < 12.00$ & 254 & $0.75 \pm 0.35$ & 91--250 yr & 563000 \\
\tableline
\tableline
\end{tabular}
\end{table}
}}


\begin{references}
\reference Cordes, J. M. \& Lazio, T. J. W. 2002, astro-ph/0207156
\reference Manchester, R. N. et al. 2001, \mnras, 328, 17 
\reference Taylor, J. H. \& Cordes, J. M. 1993, \apj, 411, 674
\reference Vivekanand, M. \& Narayan, R. 1981, J. Astrophys. Astr., 2, 315
\end{references}
\end{document}